# Dirac equation and it's squared form.

T.Khachidze[1], A. Khelashvili[2], T.Nadareishvili[3]

Department of Theoretical Physics, Ivane Javakhishvili Tbilisi State University, I.Chavchavadze ave. 3, 0128, Tbilisi, Georgia
Institute of High Energy Physics and Informatization, University str., 9, 0128, Tbilisi Georgia

**ABSTRACT** . It is shown that the squared operation of the Dirac equation which is widely applied may create new solutions and moreover may change the inner nature of original equation. Some illustrating examples are considered as well.

Dirac equation is widely used to investigate problems of relativistic quantum mechanics for fermions. There exist physically interesting cases, when the Dirac equation can be solved analytically [1]. At the same time, as a rule, the supersymmetry[2] is also appeared in these cases.

We want to notice, that squaring of Dirac equation is often used in practice to manifest a supersymmetry and to solve equation – more precisely, it reduces to ordinary second order differential equation, which can be studied by well known standard methods.

In this letter we want to pay attention to the fact, that in process of squaring of Dirac equation it is necessary to keep some caution. We imply the following: reduction of the Dirac equation to the second order equation occurs with the aid of various methods. One of the methods is manipulation on the equation, for example, expression of wave function components by the rest components and obtaining for later second order equation [3]. Other methods use squaring of the Hamiltonian or multiply on "conjugate" operator. In each cases there arise a question –get one or not some additional solutions? Below we consider typical examples and make things clear.

We consider various models of Dirac Hamiltonian and study asymptotic behavior of solutions at large distances by two methods: a) by Dirac equation and b) by its squaring. We shall see that some cases are nontrivial.

**Example 1**

Let consider following Hamiltonian

$$H = \vec{\alpha} \cdot \vec{p} + \beta m + V(r) = H_D + V(r), \qquad (1)$$

---
[1] Email: tamar_khachidze@yahoo.com
[2] Email: khelash@ictsu.tsu.edu.ge
[3] Email: t_nadar@yahoo.com

where $H_D = \vec{\alpha} \cdot \vec{p} + \beta m$ is free Dirac Hamiltonian. In this consideration $V(r)$ is zero component of Lorentz 4 – vector and it corresponds to inclusion of gauge invariant interaction in the relativistic Lagrangian. Such a Hamiltonian commutes with so called Dirac's K – operator for arbitrary central potential [4]

$$K = \beta(\vec{\Sigma} \cdot \vec{l} + 1) \quad (2)$$

Here $\vec{\Sigma}$ is spin-matrix of Dirac particle, $\vec{\Sigma} = diag(\vec{\sigma}, \vec{\sigma})$, and $\vec{l}$ is orbital momentum. It is known [5], that eigenvalues $\kappa$ of K, are degenerate with respect to 2 signs, $\pm \kappa$. It is possible to describe this degeneracy by certain A operator, which anticommutes with K [6,7]. If we demand, that this operator commutes with H, then one find, that it happens only for Coulomb potential [8]. A and AK together constitute basis of N = 2 supersymmetric algebra and give a possibility to solve algebraically the Coulomb problem [6,7,9].

Let's solve Dirac stationary equation $H\Psi = E\Psi$ for Hamiltonian (1) by standard method. Let present

$$\Psi = \begin{pmatrix} \varphi \\ \chi \end{pmatrix} \quad (3)$$

and in obtained equation exclude lower components by the relation

$$\chi = \frac{1}{E + m - V} \vec{\sigma} \cdot \vec{p} \varphi \quad (4)$$

Then for upper components one get the following equation

$$[\vec{p}^2 - (E - V)^2 + m^2]\varphi - i\frac{V'(r)}{E + m - V} \vec{\sigma} \cdot \hat{\vec{r}} \varphi = 0, \qquad \hat{\vec{r}} \equiv \frac{\vec{r}}{r} \quad (5)$$

It is obvious, that if $V(r)$ is increasing potential

$$\lim_{r \to \infty} V(r) = \infty, \quad (6)$$

then in the equation (5) leading asymptotic behaviour is

$$(p^2 - V^2)\varphi \approx 0, \quad (7)$$

or in the configuration space one get following asymptotic equation

$$\left(\frac{d^2}{dr^2} + V^2(r)\right)\varphi(r) = 0 \quad (8)$$

As $V(r)$ is increasing, for example, oscillator or linear potential, this equation have no asymptotically decreasing (normalizable) solutions, it has only oscillating solutions, i.e. this equation have no bound states (Klein paradox [10]).

Now consider equation obtained by squaring of Dirac equation, $H^2\Psi = E^2\Psi$. Calculation gives

$$\left(\vec{p}^2 + m^2 + V^2 + 2\beta mV + 2V(\vec{\alpha} \cdot \vec{p}) - i(\vec{\alpha} \cdot \hat{\vec{r}})V'(r)\right)\begin{pmatrix} \varphi \\ \chi \end{pmatrix} = E^2 \begin{pmatrix} \varphi \\ \chi \end{pmatrix} \quad (9)$$

An asymptotic form for the above class of increasing potentials is

$$((\vec{p}^2 + V^2))\begin{pmatrix} \varphi \\ \chi \end{pmatrix} \approx 0$$

We see, that one has an opposite sign, compared to (7) and therefore radial asymptotic equation is

$$\left(\frac{d^2}{dr^2} - V^2(r)\right)\varphi(r) \approx 0 \tag{10}$$

This equation has normalized solution with decreasing asymptotic, i.e. one has solutions corresponding to bound states. **So squaring of Dirac equation changes the nature of this equation.** Such a squaring remains symmetries of corresponding original equation, but there appears an additional solution, which doesn't have original Dirac equation.

So it is clear, that one should handle with care when infinitely increasing potentials are considered in the Dirac equation.

In Classical Mechanics infinitely increasing potential – isotropic oscillator gives a motion on the closed periodical orbits, but the Dirac equation for such kind of potentials have no bound states. So this classical example has no analogy in the Dirac equation, in spite of the Coulomb potential. These two potentials have different loading for Dirac equation.

**Example 2**

Let's now consider Dirac Hamiltonian with vector potentials

$$H = \vec{\alpha}\cdot\left(\vec{p} - \hat{\vec{A}}\right) + \beta m = H_D - \vec{\alpha}\cdot\hat{\vec{A}} \tag{11}$$

Here generally $\hat{\vec{A}}$ together with vector may contain also some kind of Dirac's matrices.

a) Let, $\hat{\vec{A}} = B[\vec{n}\times\vec{r}]$ - which is a vector potential of uniform magnetic field, $\vec{n}$ is the unit vector, normal to plane, which is determined by $\vec{r}$ and $\vec{p}$ vectors.

By standard manipulations on the Dirac equation, one obtains the Pauli's like equation for upper components

$$(m^2 - E^2)\varphi - (\vec{p}^2 + \vec{A}^2 + i\vec{\nabla}\cdot\vec{A} - 2\vec{A}\cdot\vec{p})\varphi - \vec{\sigma}\cdot\vec{B}\varphi - 2iB\vec{\sigma}\cdot\vec{n}(\vec{r}\cdot\vec{p})\varphi = 0 \tag{12}$$

It is obvious, that if $|\vec{A}|$ is infinitely increasing, for example $|\vec{A}| = Br$, as it is in our case, then $\vec{A}^2$ is dominant term and the problem is reduced to the asymptotic form of the oscillator potential

$$\left(\frac{d^2}{dr^2} - B^2 r^2\right)\varphi(r) \approx 0$$

i.e. Dirac equation in uniform magnetic field has bound states [1].

If we consider squared equation, one also gets right asymptotic behavior.

So in this example squared and Dirac equation have same asymptotics.

b). Let, now $\hat{\vec{A}} = im\omega\beta\vec{r}$. This is so called "Dirac oscillator" [11]. In this case Dirac equation for upper component has following form

$$(\vec{p}^2 + m^2\omega^2 r^2)\varphi - (2\vec{\Sigma}\cdot\vec{l} + 3)m\omega\varphi = (E^2 - m^2)\varphi \tag{13}$$

and equation has asymptotic behaviour, corresponding to bound states. It is interesting, that in this example by squaring we get same equation.

In conclusion, we want to underline once again that squaring of the Dirac equation may change drastically the inner nature of Dirac problems. Particularly this phenomenon occurs in cases with infinitely increasing potentials and the appropriate caution must be developed.